\documentclass{article}%
\usepackage{amsfonts}
\usepackage{amsmath}
\usepackage{amssymb}
\usepackage{graphicx}%
\providecommand{\U}[1]{\protect\rule{.1in}{.1in}}
\newtheorem{theorem}{Theorem}
\newtheorem{acknowledgement}[theorem]{Acknowledgement}

\newtheorem{definition}[theorem]{Definition}

\begin{document}

\title{Even dimensional general relativity from Born-Infeld gravity }
\author{P. K. Concha$^{1}$, D. M. Pe\~{n}afiel$^{1}$ , E. K. Rodr\'{\i}guez$^{1}$,
\ P. Salgado$^{1}$\\$^{1}$Departamento de F\'{\i}sica, Universidad de Concepci\'{o}n \\Casilla 160-C, Concepci\'{o}n, Chile}
\maketitle

\begin{abstract}
It is an accepted fact that requiring the Lovelock theory to have the maximun
possible number of degree of freedom, fixes the parameters in terms of the
gravitational and the cosmological constants. \ In odd dimensions, the
Lagrangian is a Chern-Simons form for the (A)dS group. \ In even dimensions,
the action has a Born-Infeld-like form.

Recently was shown that standard odd-dimensional General Relativity can be
obtained from Chern-Simons Gravity theory for a certain Lie algebra $B$.

Here we report on a simple model that suggests a mechanism by which standard
\ even-dimensional General Relativity may emerge as a weak coupling constant
limit of a Born-Infeld theory \ for a certain Lie subalgebra of the algebra
$B$. Possible extension to the case of even-dimensional supergravity is
briefly discussed.

\end{abstract}

\section{\textbf{Introduction}}

The most general action for the metric satisfying the criteria of general
covariance and second-order field equations for $d>4$ is a polynomial of
degree $\left[  d/2\right]  $ in the curvature known as the Lanczos-Lovelock
gravity theory $\left(  LL\right)  $ \cite{lanc},\cite{lovel}. The $LL$
lagrangian in a $d$-dimensional Riemannian manifold can be defined as a linear
combination of the dimensional continuation of all the Euler classes of
dimension $2p<d$ \cite{zum},\cite{teit}:
\begin{equation}
S=\int\sum_{p=0}^{\left[  d/2\right]  }\alpha_{p}L^{(p)} \label{uno}%
\end{equation}
where $\alpha_{p}$ are arbitrary constants and
\begin{equation}
L^{(p)}=\varepsilon_{a_{1}a_{2}\cdot\cdot\cdot\cdot\cdot\cdot a_{d}}%
R^{a_{1}a_{2}}\cdot\cdot\cdot\cdot R^{a_{2p-1}a_{2p}}e^{a_{2p+1}}\cdot
\cdot\cdot\cdot e^{a_{d}} \label{dos}%
\end{equation}
with $R^{ab}=d\omega^{ab}+\omega_{c}^{a}\omega^{cb}.$ The expression
(\ref{uno}) can be used both for even and for odd dimensions.

The large number of dimensionful constants in the $LL$ theory $\alpha_{p},$
$p=0,1,\cdot\cdot\cdot,\left[  d/2\right]  ,$ which are not fixed from first
principles, contrast with the two constants of the Einstein-Hilbert action.

In ref. \cite{tron} it was found that these parameters can be fixed in terms
of the gravitational and the cosmological constants, and that the action in
odd dimensions can be formulated as a Chern-Simons theory of the $AdS$ group.

The closest one can get to a Chern-Simons theory in even dimensions is with
the so-called Born-Infeld theories \cite{zanel} \cite{ban1}, \cite{ban2},
\cite{tron}. The Born-Infeld lagrangian is obtained by a particular choice of
the parameters in the Lovelock series, so that the lagrangian is invariant
only under local Lorentz rotations in the same way as is the Einstein-Hilbert action.

If Born-Infeld theories are the appropriate evendimensional theories to
provide a framework for the gravitational interaction, then these theories
must satisfy the correspondence principle, namely they must be related to
General Relativity. \ 

In Ref. \cite{salg1} was shown that the standard, five-dimensional General
Relativity (without a cosmological constant) can be obtained from Chern-Simons
gravity theory for a certain Lie algebra $\mathfrak{B}$.

It is the purpose of this paper to show that\ \ standard, even-dimensional
General Relativity (without a cosmological constant) emerges as a limit of a
Born-Infeld theory invariant under a certain subalgebra of the Lie algebra
$\mathfrak{B}$. On the other hand, the Lie subalgebra of the algebra
$\mathfrak{B}$, which is denoted by $\mathfrak{L}^{\mathfrak{B}}$, can be also
obtained from the Lorentz algebra and a particular semigroup $S$ by means of
the $S$-expansion procedure introduced in Refs.~\cite{salg2}, \cite{salg3}.

The Born-Infeld Lagrangian is built from the $2$-form curvature $F$\ for
$\mathfrak{L}^{\mathfrak{B}}$\ algebra which depends on a scale parameter $l$
which can be interpreted as a coupling constant that characterizes different
regimes within the theory. The field content induced by $\mathfrak{L}%
^{\mathfrak{B}}$ includes the vielbein $e^{a}$, the spin connection
$\omega^{ab}$ and one extra bosonic fields $k^{ab}$.

This paper is organized as follows: In Sec.~II we briefly review some aspect
of the Lovelock gravity theory and of the S-expansion procedure. An explicit
action for four, six and $2n$-dimensional Born-Infeld gravity invariant under
the $\mathfrak{L}^{\mathfrak{B}}$ Lie algebra is considered in Sec. III. The
weak coupling constant limit of this action is then shown to yield the
even-dimensional Einstein-Hilbert action and the corresponding Einstein field
equations. Sec.IV concludes the work with a comment about possible
developments. \ 

\section{\textbf{The Lovelock Gravity Theory}}

In this section we shall review some aspects of higher dimensional gravity and
of the S-expansion procedure. The main point of this section is to display the
differences between the invariances of $LL$ action when odd and even
dimensions are considered.

\subsection{\textbf{The local AdS Chern-Simons and Born-Infeld like\newline
gravity}}

The $LL$ action is a polynomial of degree $\left[  d/2\right]  $ in curvature,
which can be written in terms of the Riemann curvature and the vielbein
$e^{a}$ in the form (\ref{uno}), (\ref{dos}). In first order formalism the
$LL$ action is regarded as a functional of the vielbein and spin connection,
and the corresponding field equations obtained by varying with respect to
$e^{a}$ and $\omega^{ab}$ read \cite{tron}:%

\begin{equation}
\varepsilon_{a}=\sum_{p=0}^{\left[  \left(  d-1\right)  /2\right]  }\alpha
_{p}(d-2p)\varepsilon_{a}^{p}=0;\text{ \ }\varepsilon_{ab}=\sum_{p=1}^{\left[
\left(  d-1\right)  /2\right]  }\alpha_{p}p(d-2p)\varepsilon_{ab}^{p}=0
\label{tres}%
\end{equation}
where
\begin{equation}
\varepsilon_{a}^{p}:=\varepsilon_{ab_{1}\cdot\cdot\cdot b_{d-1}}R^{b_{1}b_{2}%
}\cdot\cdot\cdot R^{b_{2p-1}b_{2p}}e^{b_{2p+1}}\cdot\cdot\cdot e^{b_{d-1}}
\label{4}%
\end{equation}%
\begin{equation}
\varepsilon_{ab}^{p}:=\varepsilon_{aba_{3}\cdot\cdot\cdot a_{d}}R^{a_{3}a_{4}%
}\cdot\cdot\cdot R^{a_{2p-1}a_{2p}}T^{a_{2p+1}}e^{a_{2p+2}}\cdot\cdot
\cdot\cdot e^{a_{d}}. \label{5}%
\end{equation}
Here $T^{a}=de^{a}+\omega_{b}^{a}e^{b}$ is the torsion $2$-form. Using the
Bianchi identity one finds \cite{tron}
\begin{equation}
D\varepsilon_{a}=\sum_{p=1}^{\left[  \left(  d-1\right)  /2\right]  }%
\alpha_{p-1}(d-2p+2)(d-2p+1)e^{b}\varepsilon_{ba}^{p}. \label{siete}%
\end{equation}
Moreover
\begin{equation}
e^{b}\varepsilon_{ba}=\sum_{p=1}^{\left[  \left(  d-1\right)  /2\right]
}\alpha_{p}p(d-2p)e^{b}\varepsilon_{ba}^{p}. \label{ocho}%
\end{equation}

From (\ref{siete}) and (\ref{ocho}) one finds for $d=2n-1$%
\begin{equation}
\alpha_{p}=\alpha_{0}\frac{(2n-1)(2\gamma)^{p}}{(2n-2p-1)}\left(
\genfrac{}{}{0pt}{}{n-1}{p}%
\right)  ; \label{nueve}%
\end{equation}
with $\alpha_{0}=\frac{\kappa}{dl^{d-1}},$ $\gamma=-sign(\Lambda)\frac{l^{2}%
}{2},$ where for any dimensions $l$ is a length parameter related to the
cosmological constant by $\Lambda=\pm(d-1)(d-2)/2l^{2}$

With these coefficients, the $LL$ action is a Chern-Simons $\left(
2n-1\right)  $-form invariant not only under standard local Lorentz rotations
$\delta e^{a}=\kappa_{\text{ }b}^{a}e^{b},\quad\delta\omega^{ab}=-D\kappa
^{ab},$ but also under a local $AdS$ boost \cite{tron}.

For $d=2n$ it is necessary to write equation (\ref{siete}) in the form
\cite{tron}
\begin{equation}
D\varepsilon_{a}=T^{a}\sum_{p=1}^{\left[  n-1\right]  }2\alpha_{p-1}%
(n-p+1)\mathcal{T}_{ab}^{p}-\sum_{p=1}^{\left[  n-1\right]  }4\alpha
_{p-1}(n-p+1)(n-p)e^{b}\varepsilon_{ba}^{p} \label{diez}%
\end{equation}
with
\begin{equation}
\mathcal{T}_{ab}=\frac{\delta L}{\delta R^{ab}}=\sum_{p=1}^{\left[  \left(
d-1\right)  /2\right]  }\alpha_{p}p\mathcal{T}_{ab}^{p} \label{once}%
\end{equation}
where
\begin{equation}
\mathcal{T}_{ab}^{p}=\varepsilon_{aba_{3}\cdot\cdot\cdot\cdot a_{d}}%
R^{a_{3}a_{4}}\cdot\cdot\cdot\cdot\cdot\cdot R^{a_{2p-1}a_{2p}}T^{a_{2p+1}%
}e^{a_{2p+2}}\cdot\cdot\cdot e^{a_{d}}. \label{doce}%
\end{equation}

The comparison between (\ref{ocho}) and (\ref{diez}) leads to \cite{tron}
\begin{equation}
\alpha_{p}=\alpha_{0}(2\gamma)^{p}\binom np . \label{trece}%
\end{equation}

With these coefficients the $LL$ lagrangian takes the form \cite{tron}%
\begin{equation}
L=\frac{\kappa}{2n}\varepsilon_{a_{1}a_{2}\cdot\cdot\cdot\cdot\cdot\cdot
a_{d}}\bar{R}^{a_{1}a_{2}}\cdot\cdot\cdot\cdot\cdot\cdot\bar{R}^{a_{d-1}a_{d}}
\label{13}%
\end{equation}
which is the Pfaffian of the 2-form $\bar{R}^{ab}=R^{ab}+\frac{1}{l^{2}}%
e^{a}e^{b}$ and can be formally written as the Born-Infeld like form
\cite{tron}. The corresponding action, known as Born-Infeld action is
invariant only under local Lorentz rotations.

\subsection{\textbf{Born-Infeld Lorentz Gravity}}

A Born-Infeld action for gravity in $d=2n$ dimensions is given by \cite{tron},
\cite{zanel}%

\begin{equation}
S=\int\sum_{p=0}^{\left[  d/2\right]  }\frac{\kappa}{2n}\binom{n}{p}%
l^{2p-d+1}\varepsilon_{a_{1}\cdot\cdot\cdot\cdot a_{d}}R^{a_{1}a_{2}}%
\cdot\cdot\cdot R^{a_{2p-1}a_{2p}}e^{a_{2p+1}}\cdot\cdot\cdot e^{a_{d}}.
\label{14}%
\end{equation}
where $e^{a}$ corresponds to the 1-form \emph{vielbein}, and $R^{ab}%
=\mathrm{d}\omega^{ab}+\omega_{\ c}^{a}\omega^{cb}$ to the Riemann curvature
in the first order formalism.

The action (\ref{14}) is off-shell invariant under the Lorentz-Lie algebra
$\mathrm{SO}(2n-1,1)$, whose generators $\boldsymbol{\tilde{J}}_{ab}$ of
Lorentz transformations satisfy the commutation relationships
\[
\left[  \boldsymbol{\tilde{J}}_{ab},\boldsymbol{\tilde{J}}_{cd}\right]
=\eta_{cb}\boldsymbol{\tilde{J}}_{ad}-\eta_{ca}\boldsymbol{\tilde{J}}%
_{bd}+\eta_{db}\boldsymbol{\tilde{J}}_{ca}-\eta_{da}\boldsymbol{\tilde{J}%
}_{cb}%
\]
The Levi-Civita symbol $\varepsilon_{a_{1}...a_{2n}}$ in (\ref{14}) should be
regarded as the only non-vanishing component of the symmetric, $\mathrm{SO}%
\left(  2n-1,1\right)  ,$ invariant tensor of rank $n,$ namely
\begin{equation}
\left\langle \boldsymbol{\tilde{J}}_{a_{1}a_{2}}\cdots\boldsymbol{\tilde{J}%
}_{a_{2n-1}a_{2n}}\right\rangle =\frac{2^{n-1}}{n}\epsilon_{a_{1}\cdots
a_{2n}}.
\end{equation}

In order to interpret the gauge field as the vielbein, one is forced to
introduce a length scale $l$ in the theory. To see why this happens, consider
the following argument: Given that $(i)$ the exterior derivative operator
$\mathrm{d}=\mathrm{d}x^{\mu}\partial_{\mu}$ is dimensionless, and $(ii)$ one
always chooses Lie algebra generators $T_{A}$ to be dimensionless as well, the
one-form connection fields $\boldsymbol{A}=A_{A\mu}^{A}\boldsymbol{T}%
_{A}\mathrm{d}x^{\mu}$ must also be dimensionless. However, the vielbein
$e^{a}=e_{\text{ \ }\mu}^{a}\mathrm{d}x^{\mu}$ must have dimensions of length
if it is to be related to the spacetime metric $g_{\mu\nu}$ through the usual
equation $g_{\mu\nu}=e_{\text{ \ }\mu}^{a}e_{\text{ \ }\nu}^{b}\eta_{ab}.$
This means that the \textquotedblleft true\textquotedblright\ gauge field must
be of the form $e^{a}/l$, with $l$ a length parameter.

Therefore, following Refs.~\cite{Cha89}, \cite{Cha90}, the one-form gauge
field $\boldsymbol{A}$ of the Chern--Simons theory is given in this case by%
\begin{equation}
\boldsymbol{A}=\frac{1}{l}e^{a}\boldsymbol{\tilde{P}}_{a}+\frac{1}{2}%
\omega^{ab}\boldsymbol{\tilde{J}}_{ab}. \label{ehcs0}%
\end{equation}

It is important to notice that once the length scale $l$ is brought into the
Born-Infeld theory, the lagrangian splits into several sectors, each one of
them proportional to a different power of $l$, as we can see directly in
eq.~(\ref{14}).

\subsection{\textbf{The S-expansion procedure}}

In this subsection we shall review the main aspects of the $S$-expansion
procedure and their properties introduced in Ref. \cite{salg2}.

Let $S=\left\{  \lambda_{\alpha}\right\}  $ be an abelian semigroup with
2-selector $K_{\alpha\beta}^{\ \ \ \gamma}$ defined by
\begin{equation}
K_{\alpha\beta}^{\ \ \ \gamma}=\left\{
\begin{array}
[c]{cc}%
1 & \ \ \ \ \ \ \ \ \ \ \mbox{when}\ \ \lambda_{\alpha}\lambda_{\beta}%
=\lambda_{\gamma}\\
0 & \mbox{otherwise},
\end{array}
\right.
\end{equation}
and $\mathfrak{g}$ a Lie (super)algebra with basis $\left\{  \mathbf{T}%
_{a}\right\}  $ and structure constant $C_{AB}^{\ \ \ C}$,
\begin{equation}
\left[  \mathbf{T}_{a},\mathbf{T}_{b}\right]  =C^{\ \ \ C}_{AB}\mathbf{T}_{C}.
\end{equation}
Then it may be shown that the product $\mathfrak{G}=S\times\mathfrak{g}$ is
also a Lie (super)algebra with structure constants $C_{(A,\alpha)(B,\beta
)}^{\ \ \ \ \ \ \ \ \ \ \ \ (C,\gamma)}=K_{\alpha\beta}^{\ \ \gamma}%
C_{AB}^{\ \ \ \ C}$,
\begin{equation}
\left[  \mathbf{T}_{(A,\alpha)},\mathbf{T}_{(B,\beta)}\right]  =C^{\ \ \ C}%
_{AB}\mathbf{T}_{(C,\gamma)}.
\end{equation}
The proof is direct and may be found in Ref. \cite{salg2}.

\begin{definition}
Let $S$ be an abelian semigroup and $\mathfrak{g}$ a Lie algebra. The Lie
algebra $\mathfrak{G}$ defined by $\mathfrak{G}=S\times\mathfrak{g}$ is called
$S$-Expanded algebra of $\mathfrak{g}$.
\end{definition}

When the semigroup has a zero element $0_{S}\in S$, it plays a somewhat
peculiar role in the $S$-expanded algebra. The above considerations motivate
the following definition:

\begin{definition}
Let $S$ be an abelian semigroup with a zero element $0_{S}\in S$, and let
$\mathfrak{G}=S\times\mathfrak{g}$ be an $S$-expanded algebra. The algebra
obtained by imposing the condition $0_{S}\mathbf{T}_{A}=0$ on $\mathfrak{G}$
(or a subalgebra of it) is called $0_{S}$-reduced algebra of $\mathfrak{G}$
(or of the subalgebra).
\end{definition}

An $S$-expanded algebra has a fairly simple structure. Interestingly, there
are at least two ways of extracting smaller algebras from $S\times
\mathfrak{g}$. The first one gives rise to a \textit{resonant subalgebra},
while the second produces reduced algebras.

A useful property of the $S$-expansion procedure is that it provides us with
an invariant tensor for the $S$-expanded algebra $\mathfrak{G}=S\times
\mathfrak{g}$ in terms of an invariant tensor for $\mathfrak{g}$. As shown in
Ref. \cite{salg2} the theorem VII.2 provide a general expression for an
invariant tensor for a $0_{S}$-reduced algebra.

\textbf{Theorem VII.2 of Ref. \cite{salg2}:} \ Let $S$ be an abelian semigroup
with nonzero elements $\lambda_{i}$, $i=0,\cdots,N$ and $\lambda_{N+1}=0_{S}$.
Let $\mathfrak{g}$ be a Lie (super)algebra of basis $\left\{  \mathbf{T}%
_{A}\right\}  $, and let $\langle\mathbf{T}_{A_{n}}\cdots\mathbf{T}_{A_{n}%
}\rangle$ be an invariant tensor for $\mathfrak{g}$. The expression
\begin{equation}
\langle\mathbf{T}_{(A_{1},i_{1})}\cdots\mathbf{T}_{(A_{n},i_{n})}%
\rangle=\alpha_{j}K_{i_{a}\cdots i_{n}}^{\ \ \ \ \ j}\langle\mathbf{T}_{A_{1}%
}\cdots\mathbf{T}_{A_{n}}\rangle
\end{equation}

where $\alpha_{j}$ are arbitrary constants, corresponds to an invariant tensor
for the $0_{S}$-reduced algebra obtained from $\mathfrak{G}=S\times
\mathfrak{g}$.

\textbf{Proof:} \ the proof may be found in section $4.5$ of Ref. \cite{salg2}.

\section{\textbf{Even-dimensional Einstein-Hilbert Action \newline from
Born-Infeld gravity}}

In this section we show how to obtain the even-dimensional General Relativity
from Born-Infeld Gravity. \ 

The lagrangian for four, six-dimensional Born-Infeld gravity can be written as%

\begin{equation}
L_{\mathrm{BI}}^{\left(  4\right)  }=\frac{\kappa}{4}\varepsilon_{a_{1}%
a_{2}a_{3}a_{4}}\left(  \frac{1}{l^{4}}e^{a_{1}}e^{a_{2}}e^{a_{3}}e^{a_{4}%
}+\frac{2}{l^{2}}R^{a_{1}a_{2}}e^{a_{3}}e^{a_{4}}+R^{a_{1}a_{2}}R^{a_{3}a_{4}%
}\right)  . \label{ehcs4}%
\end{equation}%
\[
L_{\mathrm{BI}}^{(6)}=\frac{\kappa}{6}\varepsilon_{a_{1}a_{2}a_{3}a_{4}%
a_{5}a_{6}}\left(  \frac{1}{l^{6}}e^{a_{1}}e^{a_{2}}e^{a_{3}}e^{a_{4}}%
e^{a_{5}}e^{a_{6}}+\frac{3}{l^{4}}R^{a_{1}a_{2}}e^{a_{3}}e^{a_{4}}e^{a_{5}%
}e^{a_{6}}\right)
\]%
\begin{equation}
+\frac{\kappa}{6}\varepsilon_{a_{1}a_{2}a_{3}a_{4}a_{5}a_{6}}\left(  \frac
{3}{l^{2}}R^{a_{1}a_{2}}R^{a_{3}a_{4}}e^{a_{5}}e^{a_{6}}+R^{a_{1}a_{2}%
}R^{a_{3}a_{4}}R^{a_{5}a_{6}}\right)  \label{ehbi1}%
\end{equation}

From this lagrangian it is apparent that neither the $l\rightarrow\infty$ nor
the $l\rightarrow0$ limit yields the Einstein--Hilbert term alone$.$ Rescaling
$\kappa$ properly, those limits will lead either to the Euler density or to
the cosmological constant term by itself, respectively. \ In the case of large
$l$ we have\textbf{\ }$\frac{1}{l^{8}}<<\frac{1}{l^{6}}<<\frac{1}{l^{4}%
}<<\frac{1}{l^{2}}$ and in the case of small $l$ we have\textbf{\ }$\frac
{1}{l^{8}}>>\frac{1}{l^{6}}>>\frac{1}{l^{4}}>>\frac{1}{l^{2}}$. \ Since the
density of Euler is a topological invariant, that is not contribute to the
equations of motion, we have:

\textbf{(a)} \ For $d=4$ dimensions and $l\rightarrow\infty$ the dominant term
would be the Einstein Hilbert term.

\textbf{(b) }\ For dimensions $d>4$ we can see that neither the $l\rightarrow
\infty$\ nor the $l\rightarrow0$\ limit yields the Einstein--Hilbert term.

The Lagrangian (\ref{14}) is invariant under the Lorentz algebra. This algebra
choice is crucial, since it permits the interpretation of the gauge field
$\omega^{ab}$ as the spin connection. It is, however, not the only possible
choice: as we explicitly show below, there exist other Lie algebras that also
allow for a similar identification and lead to a Born-Infeld Lagrangian which
leads to the Einstein-Hilbert Lagrangian in a certain limit.

Following the definitions of Ref.~\cite{salg2} (see subsection $(2.3)$), let
us consider the $S$-expansion of the Lie algebra $\mathrm{SO}\left(
2n-1,1\right)  $ using as semigroup a sub-semigroup\ of $S_{\mathrm{E}%
}^{\left(  3\right)  }.$ After perfoming its $0_{S}$-reduction, one finds a
new Lie algebra, call it $\mathfrak{L}^{\mathfrak{B}}$ which is a subalgebra
of the so called $\mathfrak{B}$ algebra$\mathfrak{,}$with the desired properties.

\subsection{\textbf{The Lagrangian in }$\mathbf{D=4}$}

Following the definitions of Ref. \cite{salg2} (see subsection $(2.3)$), let
us consider the $S$-expansion of the Lie algebra $\mathrm{SO}\left(
3,1\right)  $ using as a semigroup the sub-semigroup $S_{0}^{\left(  3\right)
}=\left\{  \lambda_{0},\lambda_{2},\lambda_{4}\right\}  $\ of semigroup
$S_{\mathrm{E}}^{\left(  3\right)  }=\left\{  \lambda_{0},\lambda_{1}%
,\lambda_{2},\lambda_{3},\lambda_{4}\right\}  .$ After perfoming its $0_{S}%
$-reduction, one finds a new Lie algebra, call it $\mathfrak{L}^{\mathfrak{B}%
}$ which is a subalgebra of the so called $\mathfrak{B}_{5}$
algebra$\mathfrak{,}$ whose generators $J_{ab}=\lambda_{0}\tilde{J}_{ab},$
$Z_{ab}=\lambda_{2}\tilde{J}_{ab}$ satisfy the commutation relationships%

\begin{align}
\left[  J_{ab},J_{cd}\right]   &  =\eta_{cb}J_{ad}-\eta_{ca}J_{bd}+\eta
_{db}J_{ca}-\eta_{da}J_{cb},\nonumber\\
\left[  J_{ab},Z_{cd}\right]   &  =\eta_{cb}Z_{ad}-\eta_{ca}Z_{bd}+\eta
_{db}Z_{ca}-\eta_{da}Z_{cb},\\
\left[  Z_{ab},Z_{cd}\right]   &  =0,\nonumber
\end{align}
which is the expanded algebra $\mathfrak{L}^{\mathfrak{B}}=S_{0}^{\left(
3\right)  }\otimes V_{0}$. \ Using Theorem $VII.2$ of Ref.\cite{salg2} (see
subsection $(2.3)$), it is possible to show that the only non-vanishing
components of a invariant tensor for the $\mathfrak{L}^{\mathfrak{B}}$ algebra
are given by%

\begin{equation}
\left\langle J_{ab}J_{cd}\right\rangle _{\mathfrak{L}^{\mathfrak{B}}}%
=\alpha_{0}l^{2}\varepsilon_{abcd},
\end{equation}%
\begin{equation}
\left\langle J_{ab}Z_{cd}\right\rangle _{\mathfrak{L}^{\mathfrak{B}}}%
=\alpha_{2}l^{2}\varepsilon_{abcd}.
\end{equation}
where $\alpha_{0}$ and $\alpha_{2}$ \ are arbitrary independent constants of
dimensions\ $\left[  length\right]  ^{-2}$.\ \ \ 

Using the dual procedure of $S$-expansion in terms of the Maurer-Cartan forms
\cite{salg3}, we find that the Born-Infeld Lagrangian invariant under the
$\mathfrak{L}^{\mathfrak{B}}$ algebra is given by%

\begin{equation}
L_{BI\text{ \ }(4)}^{\mathfrak{L}^{\mathfrak{B}}}=\frac{\alpha_{0}}{4}%
\epsilon_{abcd}l^{2}R^{ab}R^{cd}+\frac{\alpha_{2}}{2}\epsilon_{abcd}\left(
R^{ab}e^{c}e^{d}+l^{2}D_{\omega}k^{ab}R^{cd}\right)  . \label{L4BI}%
\end{equation}

Here we can see that the Lagrangian $\left(  \text{\ref{L4BI}}\right)  $ is
split into two independent pieces, one proportional to $\alpha_{0}$ and the
other to $\alpha_{2}$. \ The term proportional to $\alpha_{0}$ corresponds to
the Euler invariant. The piece proportional to $\alpha_{2}$ contains the
Einstein-Hilbert term $\varepsilon_{abcd}R^{ab}e^{c}e^{d}$ plus a boundary
term which contains, besides the usual curvature $R^{ab},$ a bosonic matter
field $k^{ab}$.

Unlike the Born-Infeld Lagrangian $\left(  \text{\ref{ehcs4}}\right)  $ the
coupling constant $l^{2}$ does not appear explicitly in the Einstein Hilbert
term but accompanies the remaining elements of the Lagrangian. This allows
recover four dimensional the Einstein-Hilbert Lagrangian in the limit where
$l$ equals to zero.

The variation of the Lagrangian, modulo boundary terms, is given by%

\begin{equation}
\delta L_{BI\text{ \ }(4)}^{\mathfrak{L}^{\mathfrak{B}}}=\varepsilon
_{abcd}\left(  \alpha_{2}R^{ab}e^{c}\right)  \delta e^{d}+\varepsilon
_{abcd}\text{ }\delta\omega^{ab}\left(  \alpha_{2}T^{c}e^{d}+\alpha
_{2}k_{\text{ \ }e}^{c}R^{ed}\right)  .
\end{equation}
from which we see that to recover the field equations of general relativity is
not necessary to impose the limit $l=0$. \ $\delta L_{BI\text{ \ }%
(4)}^{\mathfrak{L}^{\mathfrak{B}}}=0$ leads to the dynamics of Relativity when
considering the case of a solution without matter ($k^{ab}=0$). This is
possible only in 4 dimensions. However, to recover the field equations of
general relativity in dimensions greater than 4, is necessary to take a limit
of the coupling constant $l$.

\subsection{\textbf{The Lagrangian in }$\mathbf{D=6}$}

Following the definitions of Ref. \cite{salg2} (see subsection $(2.3)$), let
us consider the $S$-expansion of the Lie algebra $\mathrm{SO}\left(
5,1\right)  $ using as a semigroup the sub-semigroup $S_{0}^{\left(  5\right)
}=\left\{  \lambda_{0},\lambda_{2},\lambda_{4},\lambda_{6}\right\}  $\ of
semigroup $S_{\mathrm{E}}^{\left(  5\right)  }=\left\{  \lambda_{0}%
,\lambda_{1},\lambda_{2},\lambda_{3},\lambda_{4},\lambda_{5},\lambda
_{6}\right\}  .$ After perfoming its $0_{S}(=\lambda_{6})$-reduction, one
finds a new Lie algebra, call it $\mathfrak{L}_{7}^{\mathfrak{B}}$ which is a
subalgebra of the so called $\mathfrak{B}_{7}$ algebra$\mathfrak{,}$ whose
generators $J_{ab}=\lambda_{0}\boldsymbol{\tilde{J}}_{ab},$ $Z_{ab}%
^{(1)}=\lambda_{2}\boldsymbol{\tilde{J}}_{ab},$ $Z_{ab}^{(2)}=\lambda
_{4}\boldsymbol{\tilde{J}}_{ab}$ satisfy the commutation relationships%

\begin{align}
\left[  J_{ab},J_{cd}\right]   &  =\eta_{cb}J_{ad}-\eta_{ca}J_{bd}+\eta
_{db}J_{ca}-\eta_{da}J_{cb},\nonumber\\
\left[  J_{ab},Z_{cd}^{\left(  1\right)  }\right]   &  =\eta_{cb}%
Z_{ad}^{\left(  1\right)  }-\eta_{ca}Z_{bd}^{\left(  1\right)  }+\eta
_{db}Z_{ca}^{\left(  1\right)  }-\eta_{da}Z_{cb}^{\left(  1\right)
},\nonumber\\
\left[  Z_{ab}^{\left(  1\right)  },Z_{cd}^{\left(  1\right)  }\right]   &
=\eta_{cb}Z_{ad}^{\left(  2\right)  }-\eta_{ca}Z_{bd}^{\left(  2\right)
}+\eta_{db}Z_{ca}^{\left(  2\right)  }-\eta_{da}Z_{cb}^{\left(  2\right)
},\nonumber\\
\left[  J_{ab},Z_{cd}^{\left(  2\right)  }\right]   &  =\eta_{cb}%
Z_{ad}^{\left(  2\right)  }-\eta_{ca}Z_{bd}^{\left(  2\right)  }+\eta
_{db}Z_{ca}^{\left(  2\right)  }-\eta_{da}Z_{cb}^{\left(  2\right)
},\nonumber\\
\left[  Z_{cd}^{\left(  2\right)  },Z_{cd}^{\left(  2\right)  }\right]   &
=\left[  Z_{cd}^{\left(  1\right)  },Z_{cd}^{\left(  2\right)  }\right]  =0.
\end{align}
which is the expanded algebra $\mathfrak{L}_{7}^{\mathfrak{B}}=S_{0}^{\left(
5\right)  }\otimes V_{0}$. \ Using Theorem $VII.2$ of Ref. \cite{salg2} (see
subsection $(2.3)$), it is possible to show that the only non-vanishing
components of a invariant tensor for the $\mathfrak{L}_{7}^{\mathfrak{B}}$
algebra are given by%
\begin{align}
\left\langle J_{ab}J_{cd}J_{ef}\right\rangle _{\mathfrak{L}^{\mathfrak{B}}}=
&  \frac{4}{3}\alpha_{0}l^{4}\varepsilon_{abcdef},\\
\left\langle J_{ab}J_{cd}Z_{ef}^{(1)}\right\rangle _{\mathfrak{L}%
^{\mathfrak{B}}}=  &  \frac{4}{3}\alpha_{2}l^{4}\varepsilon_{abcdef},\\
\left\langle J_{ab}J_{cd}Z_{ef}^{(2)}\right\rangle _{\mathfrak{L}%
^{\mathfrak{B}}}=  &  \left\langle J_{ab}Z_{cd}^{(1)}Z_{ef}^{(2)}\right\rangle
_{\mathfrak{L}^{\mathfrak{B}}}=\frac{4}{3}\alpha_{4}l^{4}\varepsilon_{abcdef}.
\end{align}
where $\alpha_{0},\alpha_{2}$ and $\alpha_{4}$ \ are arbitrary independent
constants of dimensions $\left[  \text{length}\right]  ^{-2}$.

In order to write down a Born-Infeld Lagrangian for the $\mathfrak{L}%
_{7}^{\mathfrak{B}}$ algebra, we start from the two-form curvature
\begin{align}
F  &  =\frac{1}{2}R^{ab}J_{ab}+\frac{1}{2}\left(  D_{\omega}k^{\left(
ab,1\right)  }+\frac{1}{l^{2}}e^{a}e^{b}\right)  Z_{ab}^{\left(  1\right)
}\nonumber\\
&  +\frac{1}{2}\left(  D_{\omega}k^{\left(  ab,2\right)  }+k_{\text{ \ }%
c}^{a\text{ }\left(  1\right)  }k^{cb\left(  1\right)  }+\frac{1}{l^{2}%
}\left[  e^{a}h^{\left(  b,1\right)  }+h^{\left(  a,1\right)  }e^{b}\right]
\right)  Z_{ab}^{\left(  2\right)  }.
\end{align}
which is obtained by applying the S-expansion procedure to 2-form curvature
used in the construction of the Born-Infeld action (see Appendix).

Using the dual procedure of $S$-expansion in terms of the Maurer-Cartan forms
\cite{salg3}, we find that the $6$-dimensional Born-Infeld Lagrangian
invariant under the $\mathfrak{L}_{7}^{\mathfrak{B}}$ algebra is given by%
\begin{align}
L_{BI\text{ \ }(6)}^{\mathfrak{L}^{\mathfrak{B}}}=  &  \frac{\alpha_{0}}%
{6}\varepsilon_{abcdef}l^{4}R^{ab}R^{cd}R^{ef}\nonumber\\
+  &  \frac{\alpha_{2}}{2}\epsilon_{abcdef}\left(  l^{4}\mathfrak{R}^{\left(
ab,1\right)  }R^{cd}R^{ef}+l^{2}R^{ab}R^{cd}e^{e}e^{f}\right) \nonumber\\
+  &  \frac{\alpha_{4}}{2}\varepsilon_{abcdef}\left(  l^{4}\mathfrak{R}%
^{\left(  ab,1\right)  }\mathfrak{R}^{\left(  cd,1\right)  }R^{ef}%
+l^{4}\mathfrak{R}^{\left(  ab,2\right)  }R^{cd}R^{ef}\right) \nonumber
\end{align}%
\begin{equation}
+\frac{\alpha_{4}}{2}\varepsilon_{abcdef}\left(  R^{ab}e^{c}e^{d}e^{e}%
e^{f}+l^{2}\mathfrak{R}^{\left(  ab,1\right)  }R^{cd}e^{e}e^{f}+2l^{4}%
R^{ab}R^{cd}h^{\left(  e,1\right)  }e^{f}\right)  \label{L6BI}%
\end{equation}
where
\begin{align}
\mathfrak{R}^{\left(  ab,1\right)  }  &  =D_{\omega}k^{\left(  ab,1\right)
},\\
\mathfrak{R}^{\left(  ab,2\right)  }  &  =D_{\omega}k^{\left(  ab,2\right)
}+k_{\text{ \ }c}^{a\text{ }\left(  1\right)  }k^{cb\left(  1\right)  }.
\end{align}

From $\left(  \text{\ref{L6BI}}\right)  $ we can see that in the limit $l=0$
we obtain the Einstein-Hilbert Lagrangian%
\begin{equation}
L_{BI\text{ \ }(6)}^{\mathfrak{L}^{\mathfrak{B}}}=\frac{\alpha_{4}}%
{2}\varepsilon_{abcdef}R^{ab}e^{c}e^{d}e^{e}e^{f}.
\end{equation}

Note that in the limit $l=0,$ the variation of the Lagrangian $\left(
\text{\ref{L6BI}}\right)  $ leads now to the field equations of general
relativity%
\begin{equation}
\varepsilon_{abcdef}R^{ab}e^{c}e^{d}e^{e}=0
\end{equation}%
\begin{equation}
\varepsilon_{abcdef}T^{c}e^{d}e^{e}=0.
\end{equation}

\subsection{\textbf{The Lagrangian in }$\mathbf{D=2n}$}

Following the definitions of Ref. \cite{salg2} (see subsection $(2.3)$), let
us consider the $S$-expansion of the Lie algebra $\mathrm{SO}\left(
2n-1,1\right)  $ using as a semigroup the sub-semigroup $S_{0}^{\left(
2n-1\right)  }=$ $\left\{  \lambda_{0},\lambda_{2},\lambda_{4},\lambda
_{6},\cdot\cdot\cdot,\lambda_{2n}\right\}  $\ of semigroup $S_{\mathrm{E}%
}^{\left(  2n-1\right)  }=\left\{  \lambda_{0},\lambda_{1},\lambda_{2}%
,\lambda_{3},\lambda_{4},\lambda_{5},\lambda_{6},\cdots\right.  $ $,\left.
\lambda_{2n}\right\}  .$ After perfoming its $0_{S}(=\lambda_{2n})$-reduction,
one finds a new Lie algebra, call it $\mathfrak{L}_{2n+1}^{\mathfrak{B}}$
which is a subalgebra of the so called $\mathfrak{B}_{2n+1}$
algebra$\mathfrak{,}$ whose generators $J_{ab}=\lambda_{0}\boldsymbol{\tilde
{J}}_{ab},$ $Z_{ab}^{(1)}=\lambda_{2}\boldsymbol{\tilde{J}}_{ab},$
$Z_{ab}^{(2)}=\lambda_{4}\boldsymbol{\tilde{J}}_{ab},\cdot\cdot\cdot
,Z_{ab}^{(n)}=\lambda_{2n}\boldsymbol{\tilde{J}}_{ab}$ satisfy the commutation relationships%

\begin{align}
\left[  J_{ab,}J_{cd}\right]   &  =\eta_{cb}J_{ad}-\eta_{ca}J_{bd}+\eta
_{db}J_{ca}-\eta_{da}J_{cb}\nonumber\\
\left[  J_{ab,}Z_{cd}^{\left(  i\right)  }\right]   &  =\eta_{cb}%
Z_{ad}^{\left(  i\right)  }-\eta_{ca}Z_{bd}^{\left(  i\right)  }+\eta
_{db}Z_{ca}^{\left(  i\right)  }-\eta_{da}Z_{cb}^{\left(  i\right)
}\nonumber\\
\left[  Z_{ab,}^{\left(  i\right)  }Z_{cd}^{\left(  j\right)  }\right]   &
=\eta_{cb}Z_{ad}^{\left(  i+j\right)  }-\eta_{ca}Z_{bd}^{\left(  i+j\right)
}+\eta_{db}Z_{ca}^{\left(  i+j\right)  }-\eta_{da}Z_{cb}^{\left(  i+j\right)
},
\end{align}
which is the expanded algebra $\mathfrak{L}_{2n+1}^{\mathfrak{B}}%
=S_{0}^{\left(  2n-1\right)  }\otimes V_{0}$. \ Using Theorem $VII.2$ of Ref.
\cite{salg2} (see subsection $(2.3)$), it is possible to show that the only
non-vanishing components of an invariant tensor for the $\mathfrak{L}%
_{2n+1}^{\mathfrak{B}}$ algebra are given by%

\begin{equation}
\left\langle J_{\left(  a_{1}a_{2},i_{1}\right)  }\cdots J_{\left(
a_{2p-1}a_{2p},i_{k}\right)  }\right\rangle =\frac{2^{p-1}l^{2p-2}}{p}%
\alpha_{j}\delta_{i_{1}+\cdots+i_{k}}^{j}\text{ }\varepsilon_{a_{1}\cdots
a_{2p}},
\end{equation}
where $k=0,\cdots,2n-2$ and $\alpha_{j}$ \ are arbitrary independent constants
of dimensions\ $\left[  \text{length}\right]  ^{2-2p}$.

In order to write down a Born-Infeld Lagrangian for the $\mathfrak{L}%
_{2n+1}^{\mathfrak{B}}$ algebra, we start from the two-form curvature
\begin{equation}
F=\sum_{k=0}^{n-1}\frac{1}{2}F^{\left(  ab,2k\right)  }J_{\left(
ab,2k\right)  }%
\end{equation}
where%
\begin{align}
F^{\left(  ab,2k\right)  }  &  =d\omega^{\left(  ab,2k\right)  }+\eta
_{cd}\omega^{\left(  ac,2i\right)  }\omega^{\left(  db,2j\right)  }%
\delta_{i+j}^{k}\nonumber\\
&  +\frac{1}{l^{2}}e^{\left(  a,2i+1\right)  }e^{\left(  b,2j+1\right)
}\delta_{i+j+1}^{k},\\
F^{\left(  a,2k+1\right)  }  &  =de^{\left(  a,2k+1\right)  }+\eta_{bc}%
\omega^{\left(  ab,2i\right)  }e^{\left(  c,2j\right)  }\delta_{i+j}^{k}.
\end{align}

Using the dual procedure of $S$-expansion in terms of the Maurer-Cartan forms
\cite{salg3}, we find that the $2n$-dimensional Born-Infeld Lagrangian
invariant under the $\mathfrak{L}_{2n+1}^{\mathfrak{B}}$ algebra is given by%

\begin{align}
L_{BI\text{ \ }(2n)}^{\mathfrak{L}^{\mathfrak{B}}}  &  =\sum_{k=1}^{n}%
l^{2k-2}\frac{2^{n-1}}{2n}\alpha_{j}\delta_{i_{1}+\cdots+i_{k}}^{j}%
\delta_{q_{1}+q_{2}}^{i_{1}}\cdots\delta_{q_{2n-1}+q_{2n}}^{i_{n}}\nonumber\\
&  \varepsilon_{a_{1}\cdots a_{2n}}\left(  R^{\left(  a_{1}a_{2},i_{1}\right)
}+e^{\left(  a_{1},q_{1}\right)  }e^{\left(  a_{2},q_{2}\right)  }\right)
\cdots\nonumber\\
&  \cdots\left(  R^{\left(  a_{2n-1}a_{2n},i_{k}\right)  }+e^{\left(
a_{2n-1},q_{2n-1}\right)  }e^{\left(  a_{2n},q_{2n}\right)  }\right)  .
\label{lagbi2n'}%
\end{align}

From $\left(  \text{\ref{lagbi2n'}}\right)  $ which we can see that in the
limit $l=0$ the only nonzero term corresponds to the case $k=1$, namely%

\begin{align}
\left.  L_{BI\text{ \ }(2n)}^{\mathfrak{L}^{\mathfrak{B}}}\right\vert _{l=0}
&  =\frac{2^{n-1}}{2n}\alpha_{j}\delta_{i+k_{1}+\cdots+k_{2n-2}}%
^{j}\varepsilon_{a_{1}\cdots a_{2n}}R^{\left(  a_{1}a_{2},i\right)
}e^{\left(  a_{3},k_{1}\right)  }\dots e^{\left(  a_{2n},k_{2n-1}\right)
}\nonumber\\
&  =\frac{2^{n-1}}{2n}\alpha_{j}\delta_{2p+2q_{1}+1+\cdots+2_{q_{2n-2}}+1}%
^{j}\varepsilon_{a_{1}\cdots a_{2n}}R^{\left(  a_{1}a_{2},2p\right)
}\nonumber\\
&  e^{\left(  a_{3},2q_{1}+1\right)  }\dots e^{\left(  a_{2n},2_{q_{2n-2}%
}+1\right)  }\nonumber\\
&  =\frac{2^{n-1}}{2n}\alpha_{j}\delta_{2\left(  p+q_{1}+\cdots+q_{2n-2}%
\right)  +2n-2}^{j}\varepsilon_{a_{1}\cdots a_{2n}}R^{\left(  a_{1}%
a_{2},2p\right)  }\nonumber\\
&  e^{\left(  a_{3},2q_{1}+1\right)  }\dots e^{\left(  a_{2n},2_{q_{2n-2}%
}+1\right)  }.
\end{align}
whose only nonzero component (corresponding to the case $p=q_{1}%
=\cdots=q_{2n-2}=0$) is proportional to the Einstein-Hilbert Lagrangian%

\begin{align}
\left.  L_{BI\text{ \ }(2n)}^{\mathfrak{L}^{\mathfrak{B}}}\right\vert _{l=0}
&  =\frac{2^{n-1}}{2n}\alpha_{2n-2}\varepsilon_{a_{1}\cdots a_{2n}}R^{\left(
a_{1}a_{2},0\right)  }e^{\left(  a_{3},1\right)  }\cdots e^{\left(
a_{2n},1\right)  }\nonumber\\
&  =\frac{2^{n-1}}{2n}\alpha_{2n-2}\varepsilon_{a_{1}\cdots a_{2n}}%
R^{a_{1}a_{2}}e^{a_{3}}\cdots e^{a_{2n}}.
\end{align}

\section{Comments and Possible Developments}

In the present work we have shown that standard even-dimensional general
relativity, emerges as a limit of a Born-Infeld theory invariant under a
certain algebra $\mathfrak{L}^{\mathfrak{B}}$. This algebra can be obtained
from the Lorentz algebra and a particular semigroup $S$ by means of the
$S$-expansion procedure introduced in Refs. \cite{salg2}, \cite{salg3}.

The toy model and procedure considered here could play an important role in
the context of supergravity in higher dimensions. In fact, it seems likely
that it is possible to recover the standard ten-dimensional\ Supergravity from
a Born-Infeld gravity theory, in a way very similar to the one shown here. In
this way, the procedure sketched here could provide us with valuable
information of what the underlying geometric structure of Supergravity in
$d=10$ could be (work in progress).\bigskip

\begin{acknowledgement}
\textit{This work was supported in part by FONDECYT Grants N$^{0}$ 1130653 and
by universidad de Concepci\'{o}n through DIUC Grant \ N$^{0}$ 212.011.056-1.0.
\ Three of the authors (PKC, DM, EKR) were supported by grants from the
Comisi\'{o}n Nacional de Investigaci\'{o}n Cient\'{\i}fica y Tecnol\'{o}gica
CONICYT and from the Universidad de Concepci\'{o}n, Chile.}
\end{acknowledgement}

\appendix

\section{\label{appex}$S$\textbf{-expansion of the Lorentz curvature}}

Consider the Lorentz $2$-form curvature%
\begin{equation}
\tilde{F}=\frac{1}{2}\left(  R^{ab}+\frac{1}{l^{2}}e^{a}e^{b}\right)  J_{ab},
\end{equation}
which allows to obtain the Born-Infeld action invariant under the Lorentz group%

\begin{equation}
L_{BI}^{(6)}=\frac{\kappa}{6}\epsilon_{abcdef}\left(  R^{ab}+\frac{1}{l^{2}%
}e^{a}e^{b}\right)  \left(  R^{cd}+\frac{1}{l^{2}}e^{c}e^{d}\right)  \left(
R^{ef}+\frac{1}{l^{2}}e^{e}e^{f}\right)  .
\end{equation}
The $S$-expanded $2$-form curvature is obtained as follows%
\begin{align*}
F  &  =\lambda_{\alpha}\tilde{F}^{\alpha}\\
&  =\frac{1}{2}\lambda_{0}R^{ab,0}J_{ab}+\frac{1}{2}\left(  \lambda
_{2}R^{ab,2}+\frac{1}{l^{2}}\lambda_{2}\left(  e^{a}e^{b}\right)
^{,2}\right)  J_{ab}\\
&  +\frac{1}{2}\left(  \lambda_{4}R^{ab,4}+\frac{2}{l^{2}}\lambda_{4}\left(
e^{a}e^{b}\right)  ^{,4}\right)  J_{ab}\\
&  =\frac{1}{2}R^{ab,0}J_{ab,0}+\frac{1}{2}\left(  R^{ab,2}+\frac{1}{l^{2}%
}\left(  e^{a}e^{b}\right)  ^{,2}\right)  J_{ab,2}\\
&  +\frac{1}{2}\left(  R^{ab,4}+\frac{2}{l^{2}}\left(  e^{a}e^{b}\right)
^{,4}\right)  J_{ab,4}.
\end{align*}

The Riemann curvature is given by $R^{ab}=d\omega^{ab}+\omega_{\text{ \ }%
c}^{a}\omega^{cb}$. This means that the expansion of the curvature can be
obtained by expanding the spin connection $\omega^{ab}$. For example
$R^{\left(  ab,4\right)  }$ is obtained as follows%

\[
\lambda_{4}R^{\left(  ab,4\right)  }=\lambda_{4}d\omega^{ab,4}+\lambda
_{2}\lambda_{2}\omega_{\text{ \ }c}^{a\text{ \ },2}\omega^{cb,2}+\lambda
_{4}\lambda_{0}\omega_{\text{ \ }c}^{a\text{ \ },4}\omega^{cb,0}+\lambda
_{0}\lambda_{4}\omega_{\text{ \ }c}^{a\text{ \ },0}\omega^{cb,4}.
\]
Defining $\omega^{ab,4}=k^{\left(  ab,2\right)  }$, $\omega^{ab,2}=k^{\left(
ab,1\right)  }$, $\omega^{ab,0}=\omega^{ab}$ we have%
\begin{align}
R^{\left(  ab,4\right)  }  &  =dk^{\left(  ab,2\right)  }+k_{\text{ \ }%
c}^{a\text{ \ },\left(  1\right)  }k^{\left(  cb,1\right)  }+k_{\text{ \ }%
c}^{a\text{ \ },\left(  2\right)  }\omega^{cb}+\omega_{\text{ \ }c}%
^{a}k^{\left(  cb,2\right)  }\nonumber\\
&  =dk^{\left(  ab,2\right)  }+\omega_{\text{ \ }c}^{a}k^{\left(  cb,2\right)
}+\omega^{bc}k_{\text{ \ }c}^{a\text{ \ },\left(  2\right)  }+k_{\text{ \ }%
c}^{a\text{ \ },\left(  1\right)  }k^{\left(  cb,1\right)  }\nonumber\\
&  =D_{\omega}k^{\left(  ab,2\right)  }+k_{\text{ \ }c}^{a\text{ \ },\left(
1\right)  }k^{\left(  cb,1\right)  }.
\end{align}
The field $h^{\left(  a,1\right)  }$ can be obtained from thel vielbein $e^{a}
$ as follows%
\[
\lambda_{1}\lambda_{3}e^{a,1}e^{b,3}=\lambda_{4}e^{a}h^{\left(  a,1\right)  }%
\]

So that the $2$-form curvature%
\[
F=\frac{1}{2}R^{ab,0}J_{ab,0}+\frac{1}{2}\left(  R^{ab,2}+\frac{1}{l^{2}%
}\left(  e^{a}e^{b}\right)  ^{,2}\right)  J_{ab,2}+\frac{1}{2}\left(
R^{ab,4}+\frac{2}{l^{2}}\left(  e^{a}e^{b}\right)  ^{,4}\right)  J_{ab,4}%
\]
can be rewritten as%

\begin{align}
F  &  =\frac{1}{2}R^{ab}J_{ab}+\frac{1}{2}\left(  D_{\omega}k^{\left(
ab,1\right)  }+\frac{1}{l^{2}}e^{a}e^{b}\right)  Z_{ab}^{\left(  1\right)
}\nonumber\\
&  +\frac{1}{2}\left(  D_{\omega}k^{\left(  ab,2\right)  }+k_{\text{ \ }%
c}^{a\text{ }\left(  1\right)  }k^{cb\left(  1\right)  }+\frac{1}{l^{2}%
}\left(  e^{a}h^{\left(  b,1\right)  }+h^{\left(  a,1\right)  }e^{b}\right)
\right)  Z_{ab}^{\left(  2\right)  }.
\end{align}
where $R^{\left(  ab,0\right)  }=R^{ab},$ $R^{\left(  ab,2\right)  }%
=D_{\omega}k^{\left(  ab,1\right)  },$ $\left(  e^{a}e^{b}\right)  ^{,2}%
=e^{a}e^{b},$ $R^{\left(  ab,4\right)  }=D_{\omega}k^{\left(  ab,2\right)
}+k_{\text{ \ }c}^{a\text{ }\left(  1\right)  }k^{cb\left(  1\right)  },$
$\left(  e^{a}e^{b}\right)  ^{,4}=e^{a}h^{\left(  b,1\right)  }.$

\end{document}